\documentclass[aip,sd,amsmath,amssymb,preprint,]{revtex4-1}
\usepackage{graphicx}
\usepackage{epstopdf, epsfig}
\usepackage[normalem]{ulem}
\usepackage[shortlabels]{enumitem}
\usepackage{CJK}
\usepackage{amssymb,euscript,latexsym,amsmath}
\usepackage{color}
\usepackage{epstopdf}
\usepackage{setspace}
\usepackage{chngpage}

\begin{document}

\begin{flushleft}
{\Large
\textbf\newline{Bistability in the synchronization of actuated microfilaments}
}
\newline
\\
Hanliang Guo$^1$, Lisa Fauci$^2$, Michael Shelley$^{3,4}$, and 
Eva Kanso$^{1,3}$\textsuperscript{*}

\bigskip
1. Aerospace and Mechanical Engineering, University of Southern California, Los Angeles, California, USA\\
2. Department of Mathematics, Tulane University, New Orleans, Louisiana 70118, USA\\
3. Center for Computational Biology, Flatiron Institute, Simons Foundation, New York 10010, USA\\
4. Courant Institute of Mathematical Sciences, New York University, New York, New York 10012, USA
\bigskip

* kanso@usc.edu

\end{flushleft}

\section*{abstract}
Cilia and flagella are essential building blocks for biological fluid transport and locomotion at the micron scale. They often beat in synchrony and may transition between different synchronization modes in the same cell type. Here, we investigate the behavior of elastic microfilaments, protruding from a surface and driven at their base by a configuration-dependent torque. We consider full hydrodynamic interactions among and within filaments and no slip at the surface. Isolated filaments exhibit periodic deformations, with increasing waviness and frequency as the magnitude of the driving torque increases. Two nearby but independently-driven filaments synchronize their beating in-phase or anti-phase. This synchrony arises autonomously via the interplay between hydrodynamic coupling and filament elasticity. Importantly, in-phase and anti-phase synchronization modes are bistable and co-exist for a range of driving torques and separation distances.
These findings are consistent with experimental observations of  in-phase and anti-phase synchronization \textcolor{black}{in pairs of cilia and flagella} and could have important implications on understanding the biophysical mechanisms underlying transitions between multiple synchronization modes.

\section{Introduction}

Cilia and flagella exhibit synchronous motion. The  biflagella of the alga \textit{Chlamydomonas} often beat symmetrically at the same frequency but opposite phase \cite{Ruffer1985,Ruffer1987,Polin2009,Goldstein2009,Goldstein2011}. Sperm cells tend to synchronize their tail beating in-phase when they are in close proximity~\cite{Gray1928,Woolley2009}. Motile cilia in aquatic organisms  and in mammalian tissues coordinate their collective beating in a wavelike pattern~\cite{Brennen1977, Fulford1986,Brumley2012}. 

\textcolor{black}{The origin of this synchronous behavior is attributed to mechanical coupling between the cilia, either at the cell base~\cite{Wan2016,Quaranta2015} or through hydrodynamics~\cite{Brumley2014}. In the latter,  the flagella of isolated cells exhibit synchronous beating through hydrodynamics only.} Theoretical models also suggest that synchronization can arise from hydrodynamic coupling between flagella~\cite{Mettot2011, Golestanian2011, Uchida2011, Uchida2012, Guirao2007}, assisted by flagellar elasticity \cite{Elfring2011,Goldstein2016b}. Existing models are based either on low-order representations of flagella and cilia in the form of  ``bead-spring'' oscillators~\cite{Golestanian2011, Kotar2010, Bruot2012, Niedermayer2008} or on more realistic models of hydrodynamically-coupled elastic filaments~\cite{Kim2006, Osterman2011, Goldstein2016b}.  These models  primarily reproduce one mode of synchrony: anti-phase, in-phase, or metachronal coordination. Flagellar synchrony is more complex: flagella and cilia can exhibit multiple synchronization modes even within a single cell type or organism. For example, the flagella of the algae \textit{Chlamydomonas} stochastically switch between anti-phase and in-phase synchrony \cite{Leptos2013, Wan2014}. Cilia in mammalian brain ventricles periodically change their collective beat orientation, providing a cilia-based switch for redirecting the transport of cerebrospinal fluid at regular intervals of time \cite{Faubel2016}. The origins of these transitions, whether abrupt and stochastic (\textit{Chlamydomonas} biflagellates) or gradual and periodic (ependymal cilia), are currently unknown. 

In models that represent flagella as hydrodynamically-coupled oscillators driven by a configuration-dependent force, the functional dependence of this force on configuration needs to be altered in order for the system to exhibit a different mode of synchrony~(see~\cite{Bruot2016} for review).   The need to modify the functional form of the drive, and consequently the landscape of the associated potential field, makes implicit assumptions on the mechanisms responsible for different modes of synchrony. It assumes that these mechanisms induce a fundamental change in the internal machinery that drives the flagellum or cilium beyond what can be captured by rescaling the intensity of the drive. In this study, we present a theoretical model of flow-coupled elastic filaments   that exhibits bistable in-phase and anti-phase synchronization at the same drive level, suggesting that the aforementioned  assumption is not required to achieve multiple synchronization modes.

\begin{figure}
	\centerline{\includegraphics[width=0.9\linewidth]{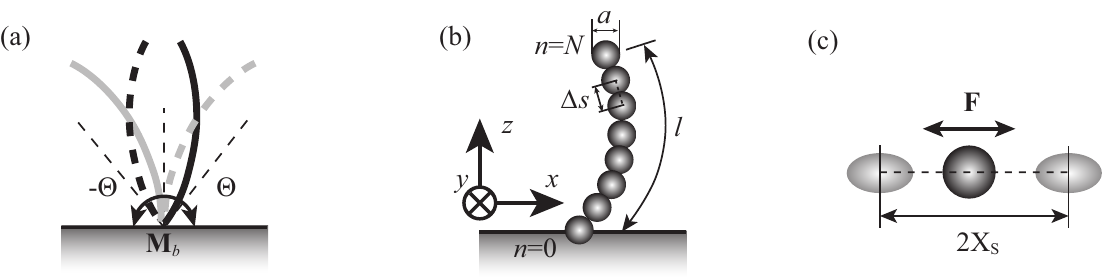}}
	\caption[]{(a) Elastic filament actuated by a motor at its base with configuration-dependent bending moment $\boldsymbol{M}_b$. The moment switches direction when the tangent to the base reaches pre-defined target angles $\pm \Theta$.  (b) Discretization of the elastic filament into $N+1$ spheres of diameter $a$.  (c) Low-order model of a shape dependent oscillator driven by an applied force $F$ that switches directions once the oscillator reaches a predefined oscillation amplitude $|x|=\mathrm{X_S}$.  }
	\label{fig:schem1}
\end{figure}

The synchronization of filaments in viscous fluid has been studied since the seminal work of \cite{Taylor1951}, where he showed that traveling waves in two parallel infinite sheets have the least viscous dissipation when synchronized in-phase. These results were later extended to include waveform compliance~\cite{Elfring2011} and three-dimensional (3D) beating~\cite{Mettot2011}.
\cite{Olson2015} considered elastic sheets and filaments of finite length and computationally showed that neighboring sheets and filaments with symmetric beating patterns always synchronize in-phase. In-phase synchrony was also predicted by \cite{Goldstein2016b}. A model of cilia that
accounted for internal actuation also demonstrated that neighboring cilia, coupled only through hydrodynamics, quickly synchronize their beat~\cite{Yang2008}.

\textcolor{black}{Cilia and flagella are driven into oscillatory motion by an intricate internal structure of microtubules and molecular motors.  Although the components of this structure are known, the mechanisms that regulate the activity of the internal motors, causing them to produce oscillatory motions are not well understood. A prominent hypothesis assumes  a geometric feedback from  mechanical deformations  to molecular activity~\cite{Brokaw1971, Brokaw2009, Riedel-Kruse2007, Sartori2016}. In its simplest form, this hypothesis supports the view that the internal forces and moments produced by the molecular motors switch ``on'' and ``off'' depending on the shape of the flagellum.}

Inspired by this switching behavior, \cite{Kotar2010} and \cite{Bruot2012} proposed optically-driven colloidal oscillators  as a model system for studying synchronization between cilia and flagella. In these systems, the colloidal particle is constrained to move on a linear trajectory under the influence of a driving force that switches direction once the particle approaches pre-defined target positions, hence the name ``geometric switch."  Here, we extend the geometric switch model to finite microfilaments submerged in viscous fluid and driven at their base by an active bending moment, of constant magnitude, that switches direction at pre-defined orientations of the tangent at the filament's base. 
\textcolor{black}{This model of internal actuation is a simplification of the biological system where actuation is applied along the filament's centerline. Yet, it carries some common features such as feedback from filament shape to internal drive.} 
\textcolor{black}{It is also reminiscent to the model used in~\cite{Kim2006} (although conceived independently), who applied a non-constant drive moment  to induce asymmetry in the filament's beat pattern for studying metachronal wave coordination of neighboring filaments.
Our study focuses on the existence of multiple modes of synchronization at constant drive magnitude.} We find that single filaments exhibit time-periodic deformations that seem to be insensitive to the initial configuration of the filament, and we quantify the consequent frequency of these deformations. We then show that two hydrodynamically-coupled filaments can achieve in-phase and anti-phase synchronizations that are bistable for a range of parameter values.
To highlight the main physical mechanisms responsible for these synchronization modes, we introduce a low-order particle model that accounts for elasticity and shape changes. The simpler model indicates that bistable synchronizations emerge as a result of hydrodynamic-coupling, shape changes and an internal restoring moment due to filament elasticity.
We conclude by commenting on the relevance of these results to undersanding the biophysical mechanisms underlying transitions between multiple synchronization modes in flagella and cilia.

\section{Continuum Model}

Consider \textcolor{black}{a nearly} inextensible elastic filament of length $\ell$ and diameter $a$ rooted at the origin $O$ of a Cartesian coordinate system $(x,y,z)$. Let $\{\boldsymbol{e}_1, \boldsymbol{e}_2, \boldsymbol{e}_3\}$ be the corresponding orthonormal basis.  The filament is free to deform in the half-space fluid domain $z\geq0$, where  $z=0$ corresponds to a no-slip solid wall (see figure~\ref{fig:schem1}). 
The centerline of the filament is denoted by the position vector $\boldsymbol{r}(s,t)$, where $s$ and $t$ represent the arc-length and time, respectively.
The balance of forces and moments on a cross section of the filament
are given by Kirchhoff's equations for an elastic  rod
~\cite{audoly2010book}
\begin{equation}\label{eq:kirchhoff}
\begin{split}
\boldsymbol{N}'-{\boldsymbol{f}}=\boldsymbol{0},\qquad
\boldsymbol{M}'+ {\hat{\boldsymbol{t}}}\times\boldsymbol{N}=\boldsymbol{0}.
\end{split}
\end{equation}
Here, the prime $(\cdot)' = \partial (\cdot)/\partial s$ denotes differentiation with respect to arc-length $s$, $\hat{\boldsymbol{t}}= {\boldsymbol{r}'}/{|\boldsymbol{r}'|}$ 
is the tangent unit vector along the filament centerline, 
$\boldsymbol{N}$ and $\boldsymbol{M}$ are the internal force and bending moment, respectively, and $-\boldsymbol{f}$ is the drag force per unit length exerted by  the surrounding fluid on the filament ($\boldsymbol{f}$ is the  force per unit length exerted by  the  filament on the  fluid). The Hookean constitutive relation between the bending moment $\boldsymbol{M}$ and the bending deformation (curvature) of the filament is given by $\boldsymbol{M}=B\hat{\boldsymbol{t}}\times\hat{\boldsymbol{t}}'$, where $B$ is the bending rigidity. The internal force $\boldsymbol{N}$ consists of a bending force and a constraint tension force that enforces the inextensibility condition. 
\textcolor{black}{The constraint is satisfied in a weak form by considering an elastic filament with large tensile stiffness.}

The filament is free at its tip $s = \ell$ and is actuated by an internal motor at its base $s=0$  that produces a torque $\boldsymbol{M}(0,t) = \boldsymbol{M}_b$. The torque $\boldsymbol{M}_b$ is a configuration-dependent torque that switches direction when the base angle $\theta_b(t)$ of the filament, defined as $\theta_b = \arcsin(\boldsymbol{e}_3\times\hat{\boldsymbol{t}}(0,t)\cdot\boldsymbol{e}_2)$, reaches predefined  target orientations $\pm \Theta$.  More specifically, we consider $\boldsymbol{M}_b=\alpha{M}_b\boldsymbol{e}_2$, where $M_b$ is a positive constant and $\alpha\in\{-1,1\}$ is a state variable that defines the torque direction;  $\alpha$ changes from $1$ to $-1$  at $\theta_b=\Theta$ and from $-1$ to $1$ at $\theta_b=-\Theta$ (see figure~\ref{fig:schem1}).
This torque model can be viewed as an extension to the geometric switch model for colloidal systems studied in~\cite{Kotar2010, Bruot2012} and as a simplified version of the ``geometric clutch'' model proposed in~\cite{Lindemann1994}.
\textcolor{black}{A target angle that acts as a geometric switch to drive elastic filaments along their entire length was used by
\cite{buchmann2017} to model the power and recovery strokes of eukaryotic cilia}.

The fluid motion is governed by the incompressible Stokes equation for zero Reynolds number flows, 
\begin{equation}\label{eq:stokes}
-\nabla p + \mu\nabla^2\boldsymbol{u} + \boldsymbol{F} = 0, \qquad \nabla\cdot \boldsymbol{u}=0.
\end{equation}
Here, $p$ is the pressure field, $\mu$ is the fluid viscosity,  $\boldsymbol{u}$ is the fluid velocity field, and $\boldsymbol{F}$ is the (Eulerian) force density exerted by the filament on the fluid.  {$\boldsymbol{F}(\boldsymbol{x},t)$ is related to the force per unit length $\boldsymbol{f}(s,t)$ as follows 
$ \boldsymbol{F}(\boldsymbol{x},t) = \int_{s\in[0,\ell]} \boldsymbol{f}(s,t) \delta(\boldsymbol{x}-\boldsymbol{r}(s,t))ds$, where
$\delta$ is the three-dimensional Dirac $\delta$-function, 
and $\boldsymbol{x}$ is the position vector.
These equations are subject to the no-slip condition $\mathbf{u} = 0$ at the bounding wall $z=0$. We take advantage of the small aspect ratio $a/\ell\ll 1 $ of the filament to approximate the velocity at the filament boundary by the velocity along its centerline, 
\begin{equation}	\label{eq:bc}
\begin{split}
\boldsymbol{u}|_{\mathrm{filament}} = 
\dot{\boldsymbol{r}}(s,t). 
\end{split}
\end{equation}

\begin{table}
\begin{center}
\begin{tabular}{lll}
\textbf{Parameter} & \textbf{Symbol} &  \textbf{Dimensional value} \\[3pt]
\hline
Filament length & $\ell$ & $20~\mu\mathrm{m}$\\
Fluid viscosity &$\mu$ & $10^{-3}~\mathrm{Pa}\cdot\mathrm{s}$ \\
Bending rigidity &$B$ & $800~\mathrm{pN}\cdot\mu\mathrm{m}^2$ \\
Time scale & $T_{\rm } = \dfrac{\ell^4 \mu}{B}$ & $0.2~$s \\
\end{tabular}
\caption{Characteristic scales of the system.}
\label{tab:parameter1} 
\end{center}
\end{table}

To fully determine the filament deformation $\boldsymbol{r}(s,t)$ given the moment $\boldsymbol{M}_b$ at the filament base, we need to solve the coupled fluid-filament system of equations~(\ref{eq:kirchhoff}--\ref{eq:bc}).  It is convenient for building an efficient numerical \textcolor{black}{method}  to (i) write the moment equation in~\eqref{eq:kirchhoff} in integral form and (ii) assume that the filament is quasi-inextensible~\cite{Teran2010, Chrispell2013, Olson2015}. In particular, we integrate the moment equation in~\eqref{eq:kirchhoff} from the filament free end at $\ell$ to any location $s$ along the filament, taking into account that $\boldsymbol{M}(\ell) = 0$ and that along the filament $ \boldsymbol{N}' = \boldsymbol{f}$. We get, after an integration by parts on the second term, that
\begin{equation}\label{eq:bforce}
\boldsymbol{M}(s)+\boldsymbol{r}\times\int_\ell^s\boldsymbol{f}\mathrm{d}\tilde{s}-\int_\ell^s\boldsymbol{r}\times\boldsymbol{f}\mathrm{d}\tilde{s}=\boldsymbol{0}.
\end{equation}
We then write the force density $\boldsymbol{f}(s,t)$ applied by the filament on the surrounding fluid as $\boldsymbol{f} = \boldsymbol{f}^{\perp} + \boldsymbol{f}^\parallel$. We assume that the force component $\boldsymbol{f}^\parallel=(\boldsymbol{f}\cdot\hat{\boldsymbol{t}})\hat{\boldsymbol{t}}$ tangent to the filament's centerline can be obtained explicitly by considering a large tensile stiffness $K$, 
\begin{equation}\label{eq:tforce}
\boldsymbol{f}^\parallel 
=-K |\boldsymbol{r}'|' \, \hat{\boldsymbol{t}},
\end{equation}
thus ensuring that the filament's length remains almost constant.
We substitute~\eqref{eq:tforce} into~\eqref{eq:bforce} taking into account that $\boldsymbol{M} = B \boldsymbol{r}'\times\boldsymbol{r}''$ to obtain an expression for $\boldsymbol{f}^\perp$ in terms of the position vector $\boldsymbol{r}(s)$ and its spatial derivatives. To this end, one gets both components of the force density $\boldsymbol{f} = \boldsymbol{f}^{\perp} + \boldsymbol{f}^\parallel$ in terms of the kinematic variables $\boldsymbol{r}$ and its derivatives. We substitute these expressions for $\boldsymbol{f}$ into~\eqref{eq:stokes} and we solve numerically subject to~\eqref{eq:bc} to obtain the nonlinear dynamics of the filament as discussed next.

To obtain non-dimensional counterparts to the equations of motion, we consider the dimensional scales associated with the fluid viscosity $\mu$ and cilium length $\ell$. Because of the geometric switch model, the system does not have an intrinsic time scale. To remedy this, we consider the time scale $ T = \ell^4 \mu /B$ arising from balancing the filament's elasticity with the fluid viscosity. To this end, we consider the bending rigidity to be of the order $B = 800~\text{pN}\cdot\mu\text{m}^2$, as reported in~\cite{Xu2016} for wild type \textit{Chlamydomonas} flagella. A list of the dimensional parameters used to scale the equations of the motion are reported in Table~\ref{tab:parameter1}. Hereafter, all quantities are considered to dimensionless unless otherwise stated.


\section{Numerical Method}
\label{sec:numerics}

We discretize the filament into a uniform chain of $N+1$ segments of length $a$ such that $\Delta s = \ell/N = a$ (figure~\ref{fig:schem1}(b)). The segments are labeled from $n=0$ at the filament base to $n=N$ at its tip. 
The position vector is discretized by $\boldsymbol{r}_n = x_n \boldsymbol{e}_1 + z_n \boldsymbol{e}_3$ and the local orientation $\theta_n$ of the tangent vector to segment $n$ is defined as the angle between the $z$-axis and the vector $\Delta \boldsymbol{r}_n = (\boldsymbol{r}_{n+1}-\boldsymbol{r}_{n})$. 
Equation~(\ref{eq:bforce}) can be written  in discrete form as follows
\begin{equation}\label{eq:momentbalance}
\boldsymbol{M}_{n-1}-\sum_{m\ge n}^N[(\boldsymbol{r}_m-\boldsymbol{r}_{n-1})\times\boldsymbol{f}_m] = \boldsymbol{0}.
\end{equation}
Here, $\boldsymbol{M}_n = - B \left[ (\theta_{n}-\theta_{n-1}) / \Delta s \right] \boldsymbol{e}_2$ for {$1\le n < N$} whereas $\boldsymbol{M}_o = \boldsymbol{M}_b$ at the no-slip wall $z = 0$.

\begin{table}
\begin{center}
\begin{tabular}{lll}
\textbf{Parameter} & \textbf{Symbol} & \textbf{Dimensionless value} \\[3pt]
\hline
Number of segments per filament &$N$ & 20 \\
Segment length &$\Delta s = a$ & $5\times10^{-2}$\\
Time step &$\Delta t$ & $2\times 10^{-5}$ \\
Total integration time &$T$ & $20$ \\
Target angle &$\Theta$ & $0.15\pi$ \\
Magnitude of base moment & $M_b$ & $0.6$ -- $3.0$\\
Filaments separation distance  & $d$ & $0.4$ -- $1.1$\\
\end{tabular}
\caption{Dimensionless parameters used in simulations ($\ell, \mu$ and $B$ are normalized to 1).}
\label{tab:parameter2} 
\end{center}
\end{table}

We decompose the force $\boldsymbol{f}_n = \boldsymbol{f}^{\perp}_n + \boldsymbol{f}^\parallel_n$ exerted by segment $n$ on the surrounding fluid into   two components $\boldsymbol{f}^{\perp}_n$ and $\boldsymbol{f}^\parallel_n$ that are perpendicular and parallel to $\Delta \boldsymbol{r}_{n-1}=(\boldsymbol{r}_n-\boldsymbol{r}_{n-1})$. We substitute into~\eqref{eq:momentbalance} and 
 rearrange the term containing $\boldsymbol{f}_n^\perp$ to the other side of the equation to get
\begin{equation}\label{eq:momentbalance2}
\Delta\boldsymbol{r}_{n-1}\times\boldsymbol{f}_n^\perp=  \boldsymbol{M}_{n-1}-\sum_{m>n}^N[(\boldsymbol{r}_m-\boldsymbol{r}_{n-1})\times\boldsymbol{f}_m].
\end{equation}
Upon taking the cross product with ${\Delta\boldsymbol{r}_{n-1}}/{\| \Delta\boldsymbol{r}_{n-1}\|^2}$, the above equation becomes
\begin{equation} \label{eq:Fb}
\begin{split}
\boldsymbol{f}_n^\perp = \left[\boldsymbol{M}_{n-1} - \sum_{m> n}^N (\boldsymbol{r}_m - \boldsymbol{r}_{n-1})\times\boldsymbol{f}_m\right]\times\frac{\Delta\boldsymbol{r}_{n-1}}{\| \Delta\boldsymbol{r}_{n-1}\|^2}.
\end{split}
\end{equation}
The parallel component $\boldsymbol{f}_n^\parallel$ is given by the discrete analog to~\eqref{eq:tforce}. Namely, 
for $1\le n <N$, one has
\begin{equation} \label{eq:Fe}
\begin{split}
\boldsymbol{f}_n^\parallel = 
&-K\left[  - \frac{\| \Delta\boldsymbol{r}_n\|-\Delta s}{\Delta s}\frac{\Delta\boldsymbol{r}_n}{\|\Delta\boldsymbol{r}_n\|}\boldsymbol{\cdot}\frac{\Delta\boldsymbol{r}_{n-1}}{\|\Delta\boldsymbol{r}_{n-1}\|} 
+   \frac{\|\Delta \boldsymbol{r}_{n-1}\|-\Delta s}{\Delta s} 
 \right]\frac{\Delta\boldsymbol{r}_{n-1}}{\|\Delta\boldsymbol{r}_{n-1}\|} ,
\end{split}
\end{equation}
whereas for $n=N$, one has
\begin{equation} \label{eq:FeN}
\boldsymbol{f}_N^\parallel = -K\frac{(\|\Delta\boldsymbol{r}_{N-1}\|-\Delta s)}{\Delta s}\frac{\Delta\boldsymbol{r}_{N-1}}{\|\Delta\boldsymbol{r}_{N-1}\|}.
\end{equation}
Using~\eqref{eq:Fb} and~(\ref{eq:Fe}-\ref{eq:FeN}),  $\boldsymbol{f}_n$ can be evaluated sequentially from the filament tip (by decreasing order of $n$) in terms of the filament kinematic variables $\boldsymbol{r}_m$, $m\geq n-1$.

\begin{figure}
        \centerline{\includegraphics[width=\linewidth]{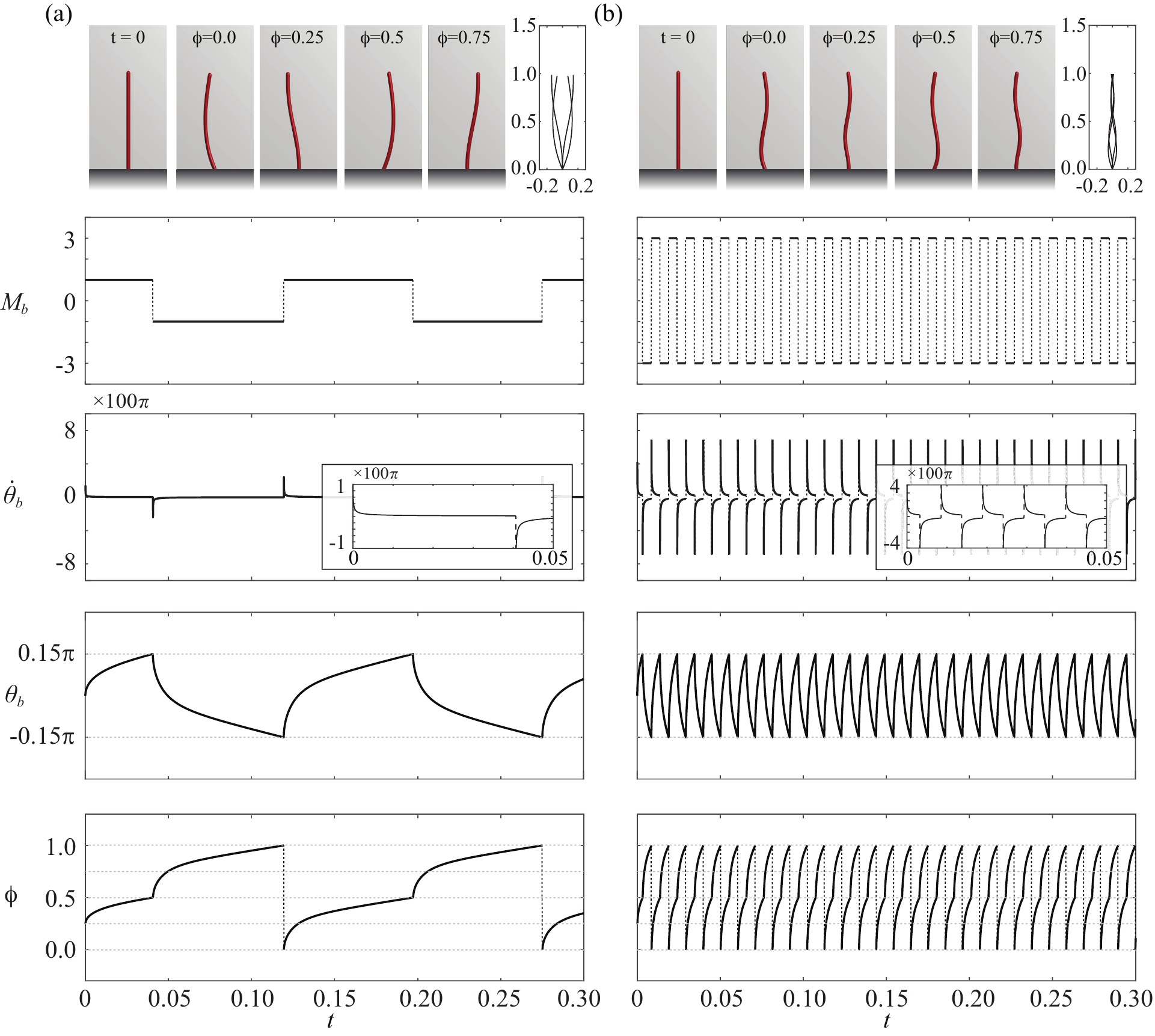}}
	\caption[]{Filament dynamics for (a) $M_b=1$ and (b) $M_b =3$. From top to bottom: 
	snapshots of the filament deformations, configuration-dependent moment $M_b$, angular velocity $\dot{\theta}_b$,  angle $\theta_b$  and phase $\phi$ at the base of the filament as functions of time. Note that the frequency of oscillations is not known a priori, only the switching angles $\Theta = \pm 0.15\pi$. 	} 
	\label{fig:single}
\end{figure}

To solve~(\ref{eq:stokes}-\ref{eq:bc}), we 
use a one-dimensional distribution of regularized Stokeslets along the centerline of the filament together with an ``image'' distribution to impose the no-slip boundary conditions at $z=0$ plane~\cite{Cortez2015}. The regularized Stokeslets are placed at the center $\boldsymbol{r}_n$ of each segment. The strength of the regularized Stokeslet at $\boldsymbol{r}_n$ is equal to the discrete force $\boldsymbol{f}_n$ and the fluid velocity generated by the filament at an arbitrary position $\boldsymbol{x}$ in the fluid domain is given by
$\boldsymbol{u}(\boldsymbol{x}) = \sum_{n=1}^N \boldsymbol{G}(\boldsymbol{x}-\boldsymbol{r}_n)\cdot\boldsymbol{f}_n$, where $\boldsymbol{G}(\boldsymbol{x}-\boldsymbol{r}_n)$ is the Green's tensor for the regularized Stokeslet near a wall~\cite{Ainley2008}.
We substitute this expression for the fluid velocity into~\eqref{eq:bc}, recalling~\eqref{eq:Fb} and~(\ref{eq:Fe}-\ref{eq:FeN}) to express $\boldsymbol{f}_n$ in terms of the filament position $\boldsymbol{r}_m$. This yields a set of coupled equations for the filament dynamics that we evolve forward in time using the forward Euler \textcolor{black}{method}. Initial conditions for this system are the configuration of the filament $\mathbf{r}(s,t)$ and the state variable $\alpha = \pm 1$.

In all numerical simulations, we fix the target angles at $\Theta=0.15\pi$, we use $N = 20$ segments of length $a = 1/20$  to discretize the filament, and we set the regularization parameter of the Stokeslet to be equal to $a$. 
The tensile stiffness $K=5000$ is a numerical parameter chosen to keep the length of the filament almost constant. 
The filaments are initialized in straight configuration normal to the wall.
Here, the time step is $0.2\times 10^{-4}$ and the system is integrated to $T=20$, which is sufficient to capture the long-time dynamics; see Table~\ref{tab:parameter2} for a summary of all parameter values.

\section{Deformation of a Single Filament}

\begin{figure}
	\centerline{\includegraphics{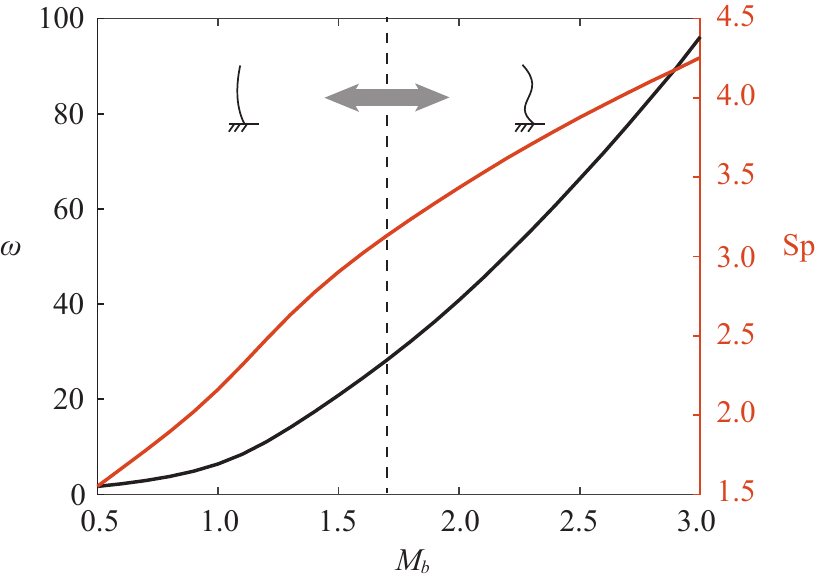}}
	\caption[]{
	Beating frequency $\omega$ and Sperm number $\mathrm{Sp}$ of a single filament as a function of bending moment at the base $M_b$. }
	\label{fig:single_freq}
\end{figure}

We examine the motion of a single filament actively driven at its base by \textcolor{black}{$\boldsymbol{M}_b = \alpha M_b \boldsymbol{e}_2$ ($\alpha = \pm 1$) that switches direction at  $\Theta=\pm 0.15\pi$. Figure~\ref{fig:single} shows the deformations and time evolution} of the filament for two distinct values of the bending moment: (a) $M_b = 1$ and (b) $M_b = 3$. 
In each case, the moment $M_b$ is constant in magnitude between switches (figure~\ref{fig:single}, \textcolor{black}{second} row). In response, the angular velocity $\dot{\theta}_b$ slows down after each switching event \textcolor{black}{(figure~\ref{fig:single}, third row)}. This decrease in $\dot{\theta}_b$ is induced by the internal restoring moment due to the filament elasticity, which acts in the opposite direction to $\boldsymbol{M}_b$.  

We introduce the phase variable $\phi \in [0,1]$ as a linear interpolation of $\theta_b$ between the two target angles $\pm\Theta$,
\begin{equation}
\label{eq:phase}
\begin{split}
\phi =  \frac{\Theta+\alpha \theta_b}{4\Theta}+\frac{1-\alpha}{4} =  \left\{ \begin{array}{l l}
\dfrac{\Theta+\theta_b}{4\Theta}, &  \alpha=1, \\[4ex]
\dfrac{3\Theta-\theta_b}{4\Theta}, &  \alpha = -1.
\end{array} \right. 
\end{split}
\end{equation}
By definition, the values of $\phi$ lie in $ [0, 0.5]$ when the base moment is positive ($\alpha=1$) and $\phi \in[0.5,1]$ when the base moment switches to negative  ($\alpha=-1$). Thus, the phase variable $\phi$ is monotonic in time over one oscillation period (see figure~\ref{fig:single}, bottom row) and as a result, it can be viewed as a time re-parameterization for examining the long-term periodic behavior of the filament. We therefore label the snapshots in figure~\ref{fig:single}, top row, by their phase $\phi$.

\begin{figure}
	\centerline{\includegraphics[width=\linewidth]{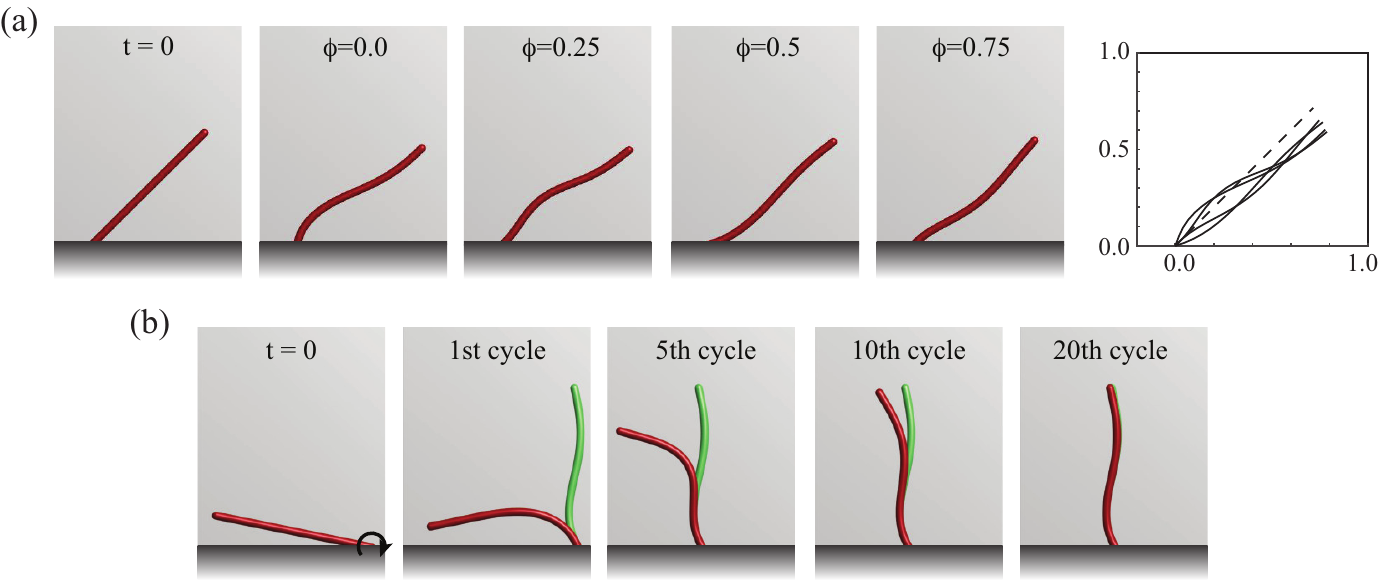}}
	\caption[]{
	(a) Filament with tilted angles produce asymmetric beating patterns. $M_b=3$, target angles are $\pi/4\pm\Theta$. Dashed line shows the average orientation $\pi/4$.
	(b) The filament with $M_b=3$ recovers the same beating pattern when starting from a perturbed initial condition; the long-term beating pattern shown in green for comparison. 
	\textcolor{black}{Filament with lower base moment converge to the long-term beating pattern within fewer cycles (see supplementary movies 1-4).
	}} 	\label{fig:single_snaps}
\end{figure}

The frequency of switching in the drive is a property of the system that depends on the system's parameters, including the magnitude of the base moment $M_b$. Figure~\ref{fig:single} indicates that for $M_b=1$, the base angle $\theta_b$ takes longer time to reach $\pm\Theta$ than for $M_b = 3$. To quantify \textcolor{black}{the} frequency $\omega$ of switching, we define it as the number of left-side switching events per unit time, averaged over late-time behavior.  Figure~\ref{fig:single_freq} depicts  $\omega$ as a function of $M_b$.  The frequency $\omega$ increases monotonically with $M_b$ as well as the waviness of the filament.  For $M_b$ larger than  $1.7$, the curvature along the filament $(0< s < \ell)$ changes sign at least once, indicating a clear traveling-wave pattern.
The switching frequency can be mapped to a Sperm number $\mathrm{Sp}=(\ell^4\xi_\perp\omega/B)^{1/4}$, defined as the ratio between the filament length and the elasto-viscous penetration length~\cite{Wiggins1998,Lagomarsino2003, Eloy2012}. Here,
$\xi_\perp = 4\pi\mu/\ln(2\ell/a)$ is the local drag coefficient perpendicular to the filament direction~\cite{Lauga2009}. The Sperm number Sp increases monotonically with $M_b$, ranging from $1.6$ to $4.3$ for $M_b \in [0.5,3]$. 
This range of Sperm numbers is consistent with those observed empirically in flagella and cilia. For example, the  wild type \textit{Chlamydomonas} has a beating frequency of 60-80~Hz~\cite{Leptos2013}, which yields a Sperm number $\mathrm{Sp}\approx 3$, given the characteristic parameters listed in Table~\ref{tab:parameter1}.

\textcolor{black}{Although the Sperm number is comparable to that of cilia and flagella, the filament deformations deviate from those observed in nature in that the amplitude of the traveling-wave decreases towards the tip. This is due to the fact that the model considers a driving moment at the base only, while in many biological filaments, the moments are distributed along the whole filament.
In the model, the beating pattern is related to the choice of the target angle $\Theta$; larger $\Theta$ produces beating patterns with high curvature.} Further, the filament deformations are symmetric because the switching orientations $\pm\Theta$ are equal and opposite. To break this symmetry, it suffices to tilt the angle about which the geometric switch is applied  by setting $\Theta_{\rm left} = 0.1\pi$ and $\Theta_{\rm right} = 0.4\pi$ as shown in figure~\ref{fig:single_snaps}(a). Hereafter, we restrict our discussion to the symmetric case.
Finally, we note that  the  long-term behavior of the filament depends on $M_b$ but it is independent of the filament initial configuration, as illustrated in figure~\ref{fig:single_snaps}(b).


\section{Synchronization of Two Filaments}

We consider the behavior of two hydrodynamically-coupled, elastic filaments separated by a distance $d$ and subject to the same moment $\boldsymbol{M}_b$ at their base.
\textcolor{black}{We set the separation distance to be large enough ($d>0.4$) so that the two filaments do not intersect.}
\begin{figure}
	\centerline{\includegraphics[width=\linewidth]{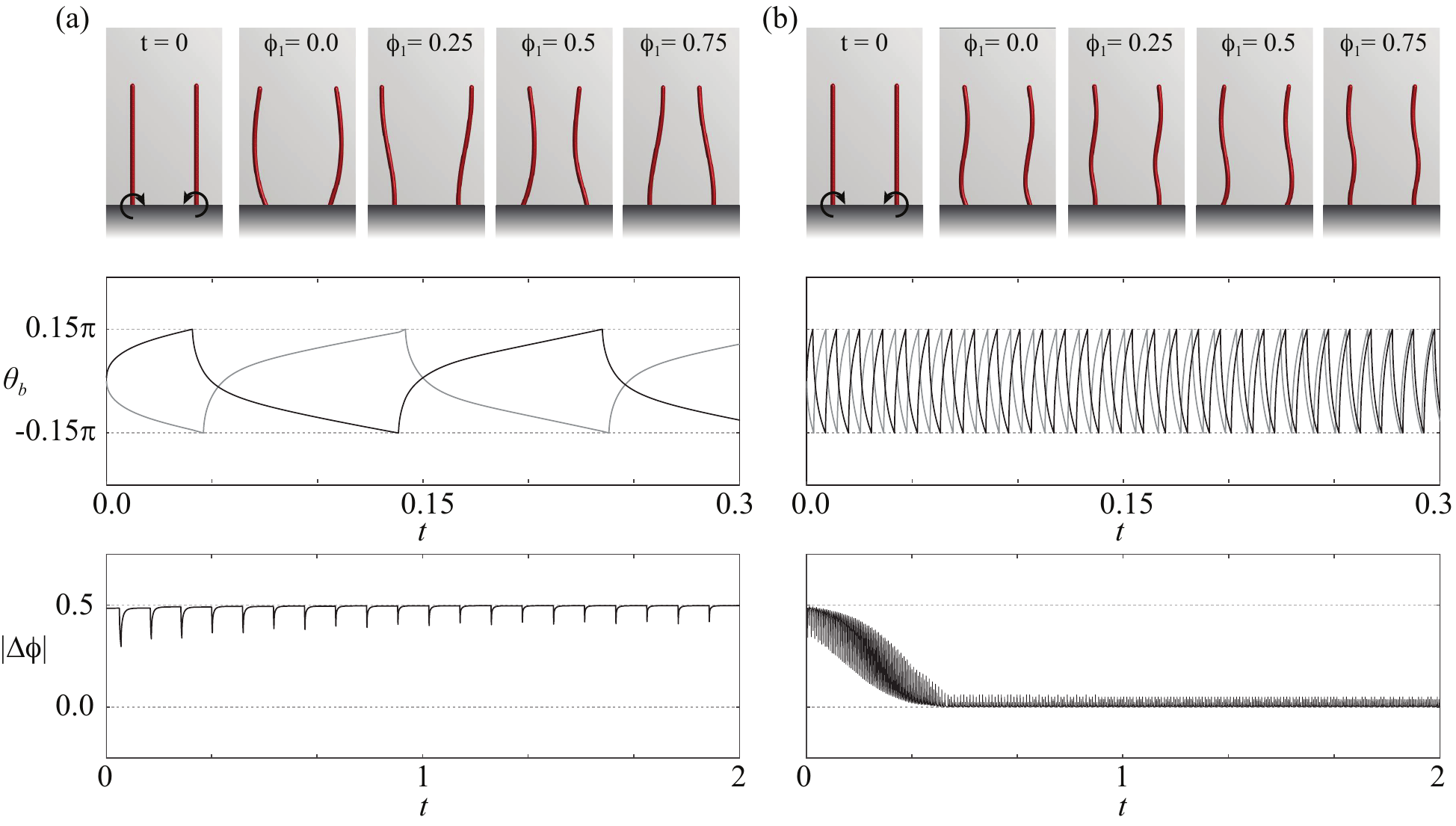}}
	\caption[]{Long term dynamics of \textcolor{black}{a pair of} filaments with (a) $M_b=1$ and (b) $M_b=3$ for $d=0.7$ and $\Delta\phi(0)=0.49$.
	Top: beating patterns at four different phases ($\phi_1=0,0.25,0.5,0.75$). 
	Middle: base angles $\theta_b$ as a function of time for $0<t<0.3$ (to highlight transient behavior). 
	Bottom: phase difference $|\Delta\phi|$ for $0<t<2$. 
	}
	\label{fig:double_snaps}
\end{figure}

%
Figure~\ref{fig:double_snaps} shows the long-term behavior of two filaments that start in a nearly anti-phase configuration; the left filament, referred to as filament 1, is initially straight and moving to the right ($\alpha_1(0) = 1$) whereas the right filament, or filament 2, is initially moving to the left ($\alpha_2(0) = -1$) such that the phase difference is equal to $\Delta \phi(0) = \phi_2(0) - \phi_1(0)= 0.49$. Here, the state variables $\alpha_1$ and $\alpha_2$ and phase variables $\phi_1$ and $\phi_2$ for filaments 1 and 2 are defined as in~\eqref{eq:phase}. This initial configuration corresponds to a small perturbation away from the anti-phase configuration for which $\Delta \phi = 0.5$.
 The coupling between the two filaments is due to hydrodynamic interactions only.

The two filaments exhibit anti-phase synchronization for $M_b = 1$ and $d=0.7$ as shown in figure~\ref{fig:double_snaps}(a),  whereas for $M_b = 3$ the two filaments depart from their anti-phase initial conditions and approach in-phase synchronization as shown in figure~\ref{fig:double_snaps}(b).
\textcolor{black}{These  modes of synchronization are quantified in figure~\ref{fig:double_snaps} \textcolor{black}{bottom row}. }
In both cases, the shapes of the filaments show no significant difference compared to those exhibited by the single filaments.

To quantify the long-term synchronization mode between the two filaments, we adapt the synchronization order parameter ${Q}$  proposed in \cite{Kotar2010}. Namely, we let
\begin{equation}\label{eq:syncorder}
Q=\frac{-1}{T-T^*}\int_{T^*}^{T} \alpha_1(t)\alpha_2(t)\mathrm{d}t,
\end{equation}
where $T$ is the total integration time, $T^*$ is chosen to ensure that transient behavior is excluded. By construction, one has $-1\le Q \le 1$, where $Q=-1$ describes exactly in-phase motions while $Q=1$ corresponds to exactly anti-phase motions. In the simulations, we set $T^*=15$ and $T=20$ time units, respectively; the filaments are said to be in-phase if \textcolor{black}{$Q\in [-1,-0.5]$} and anti-phase if \textcolor{black}{$Q\in [0.5,1]$}.

\begin{figure}
	\centerline{\includegraphics[width=0.9\linewidth]{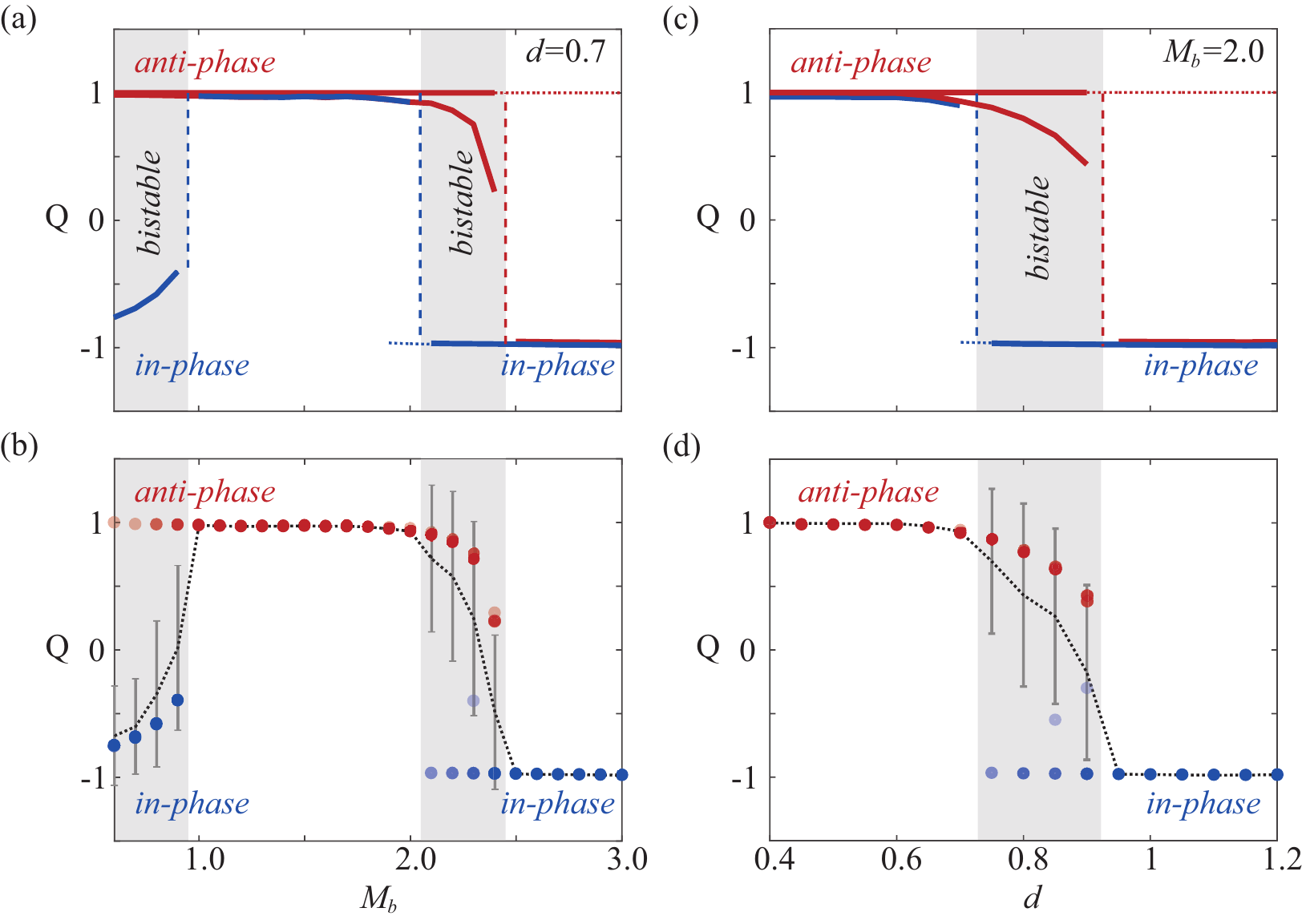}}
	\caption[]{\textcolor{black}{Synchronization order parameter $Q$ as a function of bending moment $M_b$ (Panels (a) and (b)), and as a function of separation distance $d$ (Panels (c) and (d)). 
	(a) and (c) In-phase initial conditions are shown in blue and anti-phase in red. (b) and (d) Mean $\langle Q \rangle$ (dotted line) and standard deviation $\mathrm{SD}(Q)$ (grey error bars) corresponding to 20 random initial conditions; $Q$ values are shown as blue and red dots, with color intensity proportional to the percentage of initial conditions resulting in these values. 	}
	}
	\label{fig:double_Q}
\end{figure}

\begin{figure}
	\centerline{\includegraphics[width=0.9\linewidth]{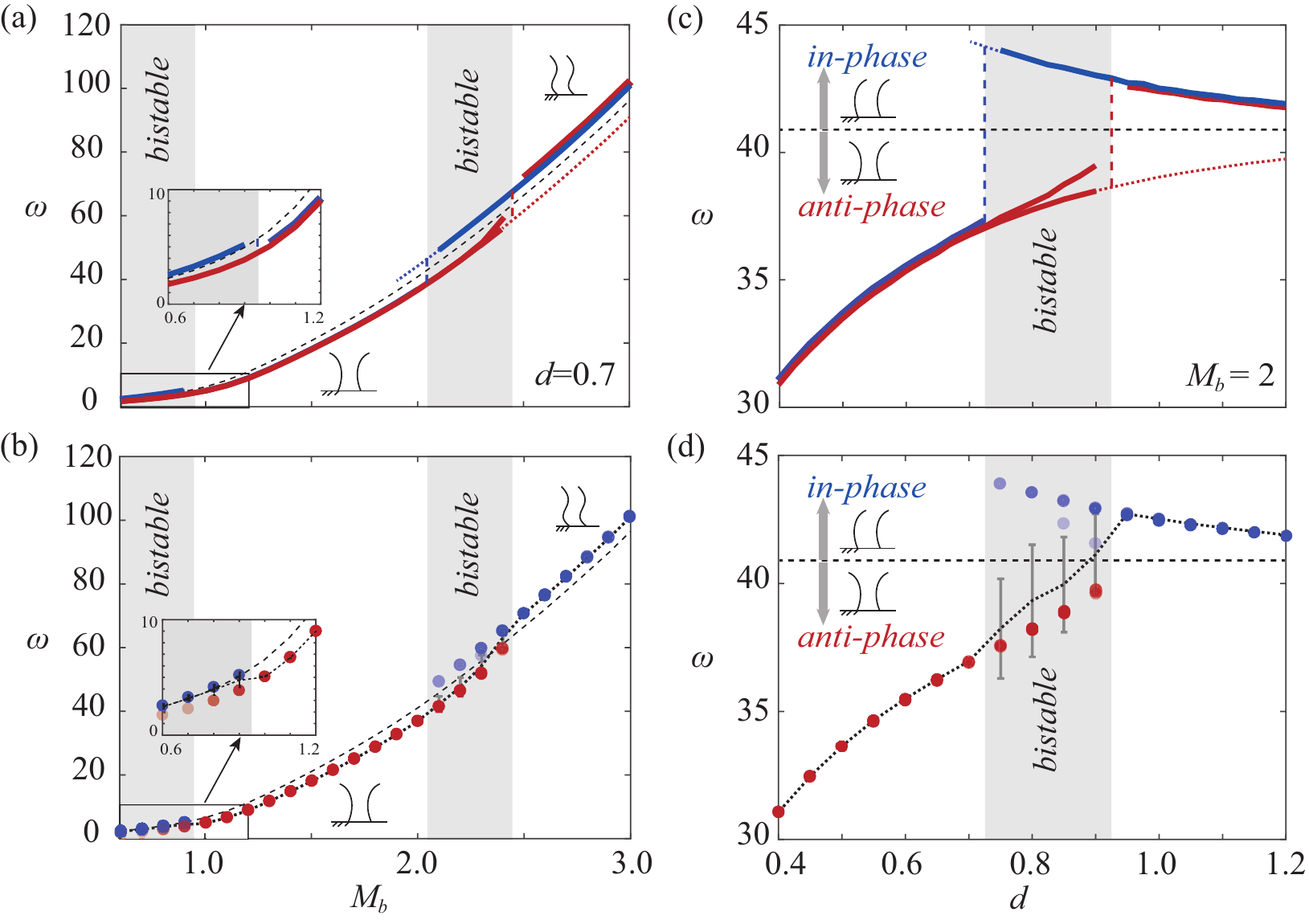}}
	\caption[]{
\textcolor{black}{Beating frequency $\omega$ as a function of bending moment $M_b$ (Panels (a) and (b)), and as a function of separation distance $d$ (Panels (c) and (d)). 
	(a) and (c) In-phase initial conditions are shown in blue and anti-phase in red. (b) and (d) Mean $\langle \omega \rangle$ (dotted line) and standard deviation $\mathrm{SD}(\omega)$ (grey error bars) corresponding to 20 random initial conditions; $\omega$ values are shown as blue and red dots, with color intensity proportional to the percentage of initial conditions resulting in these values. In all panels, the frequency of a single filament  (dashed black line) is superimposed  for comparion.	}
	 }
	\label{fig:double_freq}
\end{figure}

We fix the separation distance between the filaments at $d=0.7$ and investigate the effect of the bending moment $M_b$ on the long-term synchronization modes between the two filaments. We consider in-phase and anti-phase initial conditions $\Delta\phi(0) = 0$ and $\Delta\phi(0) = 0.5$, respectively, as well as small perturbations $\Delta\phi(0)=0.01$ and $\Delta\phi(0)=0.49$ away from these configurations. Figure~\ref{fig:double_Q}(a) depicts the synchronization order parameter $Q$  versus $M_b$ shown in blue for $\Delta\phi(0) = 0$ and $\Delta\phi(0) = 0.01$ and  in red  for $\Delta\phi(0) = 0.5$ and $\Delta\phi(0) = 0.49$. When starting at  $\Delta \phi = 0.5$, the filaments always synchronize anti-phase. However, this anti-phase synchronization become unstable for large $M_b$ (dashed red line) because the filaments shift to in-phase synchronization under a small perturbation in the initial conditions ($\Delta\phi(0) = 0.49$). On the other hand, when starting at $\Delta \phi = 0$ and $\Delta \phi = 0.01$, the filament synchronize in-phase for small $M_b$, shift to anti-phase synchronization as $M_b$ increases, and shift back to in-phase synchronization as $M_b$ increases further. For $0.6\le M_b\le0.9$ and $2.1\le M_b\le2.4$, the filaments exhibit both stable in-phase and stable anti-phase synchronization depending on initial conditions. To better understand the sensitivity of these synchronization modes to perturbations in the initial conditions, we perform Monte Carlo simulations with initial conditions randomly chosen from a uniform distribution function $\Delta\phi(0)\in\mathcal{U}(-0.5,0.5)$. 
Statistical results of the synchronization modes based on 20 Monte Carlo simulations are shown in figure~\ref{fig:double_Q}(b). Dotted lines and error bars depict the mean ${\langle Q \rangle}$ and standard deviation SD$(Q)$ of the synchronization order parameter respectively. Overlaid blue and red dots are the distributions of the Monte Carlo simulations, colored in blue and red according to the emergent synchronization modes (blue for in-phase and vice-versa). The color intensity of the dots represents the fraction of simulations corresponding to a particular $Q$ value -- lighter color means fewer simulations out of 20 total number of simulations. The bistable regions where both red and blue dots co-exist are consistent with the results in figure~\ref{fig:double_Q}(a). In the bistable regions,  the synchronization mode is sensitive to initial conditions. 

To explore the effect of the separation distance $d$ between the filaments on the emergent synchronization modes, we fix the magnitude of the bending moment at $M_b=2$ and plot the synchronization order parameter $Q$ versus $d$ in figure~\ref{fig:double_Q}(c,d). For small $d$, all initial conditions lead to anti-phase synchronization. As $d$ increases, both in-phase and anti-phase synchronizations co-exist, depending on initial conditions, and as $d$ increases further, only in-phase synchronization are observed. The Monte Carlo simulations shown in figure~\ref{fig:double_Q}(d) are consistent with these findings.
\textcolor{black}{In the limit $d\rightarrow\infty$, the two filaments maintain their initial phase difference. In other words, as $d\rightarrow\infty$, the two filaments will take infinitely long time to synchronize.}

In Figure~\ref{fig:double_freq}, we report the values of the emergent beating frequencies for the cases considered in figure~\ref{fig:double_Q} and compare these values to the case of a single filament from figure~\ref{fig:single_freq}, which we show in black dashed lines in figure~\ref{fig:double_freq}.
The beating frequencies for the pair of filaments are either faster or slower than the single filament depending on their synchronization modes: anti-phase filaments beat at lower frequencies because the two filaments ``work against each other'' while in-phase filaments beat at higher frequencies  because they ``work together''. 
In particular, in anti-phase beating the two fibers are compressing and extending fluid elements in the region between them while for in-phase beating, the fluid and the filaments move together.
\textcolor{black}{In the limit $d\rightarrow\infty$, the beating frequency for the pair of filaments converges to the beating frequency of a single filament.}

Figure~\ref{fig:overlay} shows the synchronization order parameter $Q$ over the parameter space $(d,M_b)$; figure~\ref{fig:overlay}(a) shows $Q$ for nearly in-phase initial conditions $\Delta\phi(0)=0.01$ and figure~\ref{fig:overlay}(b)  for nearly anti-phase initial conditions $\Delta\phi = 0.49$. Figures~\ref{fig:overlay}(c,d) show the mean and standard deviation, respectively, of $Q$ for 20 Monte Carlo simulations with initial phase differences chosen from a uniform distribution $\mathcal{U}(-0.5,0.5)$. Taken together, these results imply that the parameter space can be divided into three distinct regions: a stable anti-phase region where \textcolor{black}{$\langle Q\rangle \in [0.5,1]$} and $\mathrm{SD}(Q)<0.2$; a stable in-phase region where  \textcolor{black}{$\langle Q\rangle \in [-1,-0.5]$} and $\mathrm{SD}(Q)<0.2$; and a bistable region where \textcolor{black}{$\langle Q \rangle \in [-0.5, 0.5]$} and $\mathrm{SD}(Q)>0.2$, in which the synchronization states are sensitive to the initial phase differences. The three regions are illustrated in figure~\ref{fig:overlay}(e).

\begin{figure}
\centerline{\includegraphics[width=\linewidth]{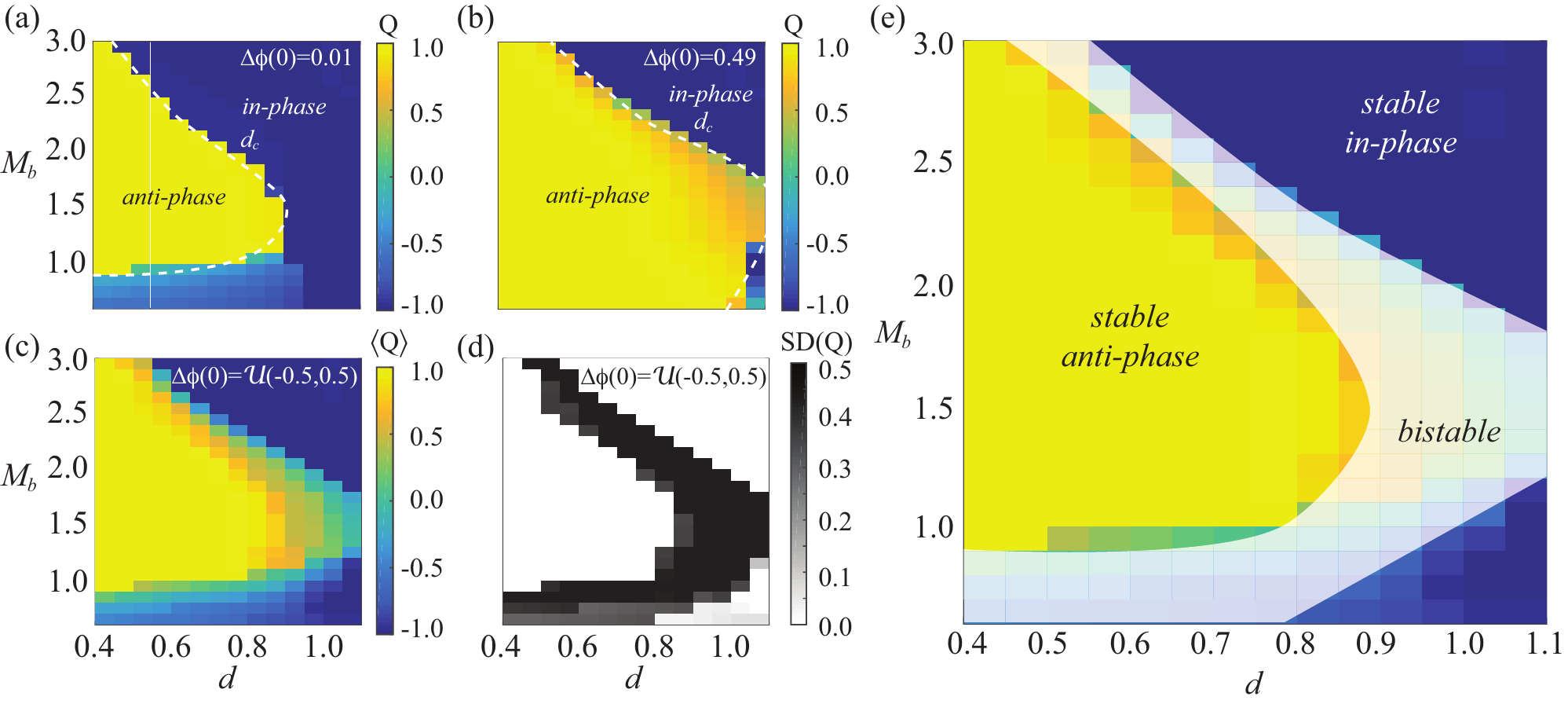}}
	\caption[]{Synchronization order parameter of two filaments as a function of bending moment $M_b$ and separation distance $d$ for
	(a) nearly in-phase ($\Delta\phi(0)=0.01$), (b) nearly anti-phase ($\Delta\phi(0)=0.49$), and (c) and (d)  20 randomly chosen initial conditions. $\langle Q\rangle$ and  $\mathrm{SD}(Q)$ are the mean and standard deviation of synchronization order parameter.  High $\mathrm{SD}(Q)$ indicates sensitivity to initial conditions.
	(e)  In-phase, anti-phase and bistable regions obtained by overlaying (c) and (d).
	}
	\label{fig:overlay}
\end{figure}

\section{Shape-dependent Oscillators}

In the geometric-switch model proposed by~\cite{Kotar2010} and \cite{Bruot2012}, a rigid spherical particle is free to move along one-direction, say the $x$-axis, under the influence of a driving force  $F$ that switches direction when the particle position reaches predefined target positions. Here, we develop a phenomenological model, based on the geometric switch oscillator, that accounts for the filament's elasticity and shape changes in terms of a ``lumped" shape variable, which we denote by $ \mathrm{s}$ (not to be confused with the filament's arclength $s$);  see figure~\ref{fig:schem1}(c).
We propose the coupled position-shape system of equations 
\begin{equation}\label{eq:simp}
\begin{split}
\xi(\mathrm{s})\dot{x}  & = -k\mathrm{s} + \alpha F, \\[0.5ex]
\tau \dot{\mathrm{s}} & = -\mathrm{s} + \alpha F.
\end{split}
\end{equation}
The  elastic ``particle" is subject to  a configuration-dependent force  $\alpha F$, where the magnitude $F$ is constant whereas $\alpha$ switches between $\{-1,1\}$ as the particle position reaches a predefined oscillation amplitude $|x|= \mathrm{{X}_s}$.  
For a rigid particle, the drive $F$ is balanced by a hydrodynamic drag equal to $\xi \dot{x}$, where $\xi$ is a constant (positive) drag coefficient. Elasticity introduces an internal restoring force that competes with the driving force and couples the shape $\mathrm{s}$ to the orientation dynamics. The elastic force is modeled via a spring with stiffness coefficient $k$ that represents a ``lumped" elastic modulus of the filament. 
The drag coefficient $\xi$ also depends on shape; it should be maximum when $\mathrm{s}=0$, that is, when the filament is straight and moving transversally to itself, and minimum when the filament reaches its maximum deformation. It should also be  symmetric under reflections from $\mathrm{s}$ to $-\mathrm{s}$. 
We therefore set $\xi(\mathrm{s})= \max(\xi_o - b\mathrm{s}^2,\epsilon)$, where the quadratic function $\xi_o - b\mathrm{s}^2$ is maximum at and symmetric about $\mathrm{s} = 0$ and the parameter $b$ characterizes the  dependence of $\xi$ on shape. The lower bound $\epsilon>0$ ensures that the drag coefficient $\xi$ remains positive at all time.

The shape of the filament changes under the influence of the driving force but relaxes to its original shape when it is not actuated. In~\eqref{eq:simp}, we assume that $\alpha F$ drives the shape directly and that the shape $\mathrm{s}$ relaxes to the original shape $\mathrm{s}_o$ with constant relaxation parameter $\tau$. For a fixed value of $\alpha$, the solution to the shape equation is of the form $\mathrm{s} = \alpha F + (\mathrm{s}_o - \alpha F) e^{-t / \tau}$, where  $\mathrm{s}$ relaxes to $\mathrm{s}_o$ when $F=0$. For non-zero $F$, the force switches sign at $\pm \mathrm{X_s}$, thus coupling position and shape. An alternative form of the shape equation in~\eqref{eq:simp} could be written by using $\dot{x}$ to drive the shape dynamics instead of directly driving it by $\alpha F$. Then, the shape equation becomes nonlinear. We chose the linear form in~\eqref{eq:simp} because we are mainly interested in reducing the complexity of the dynamical system, while identifying the main physical mechanisms at play.

\begin{figure}
	\centerline{\includegraphics[width=\linewidth]{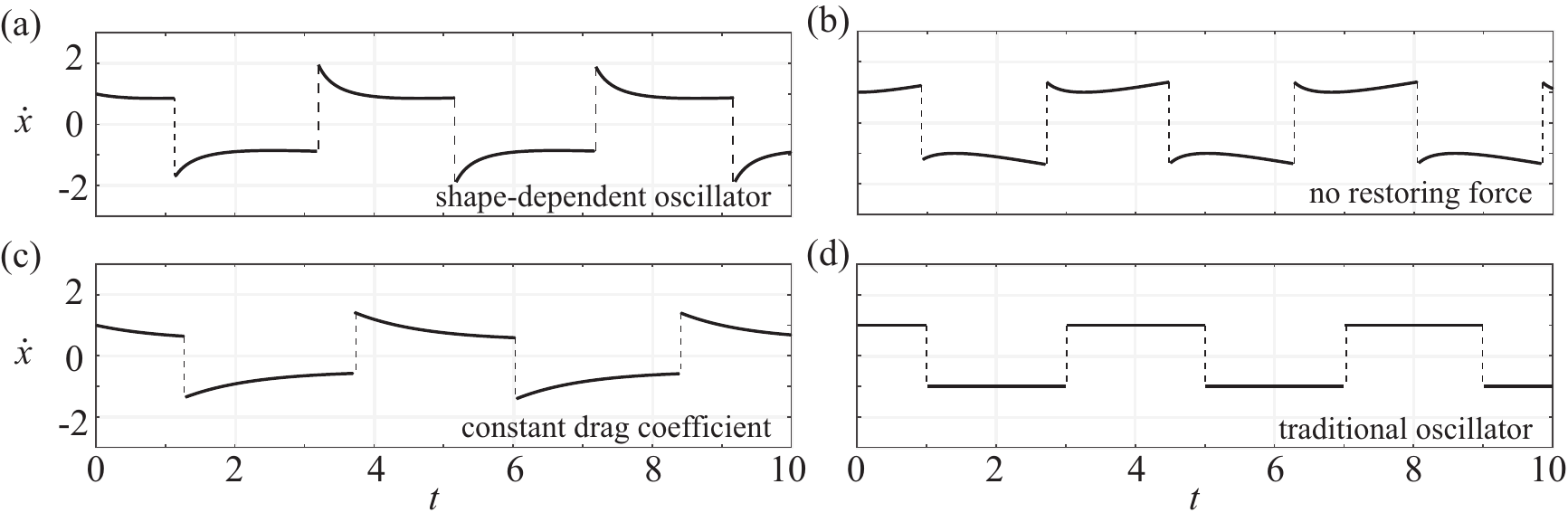}}

	\caption[]{Velocity evolution $\dot{x}$ versus time $t$ of the single shape-dependent oscillator for: (a) Shape-dependent oscillator, $b=0.5$, $k=0.5$; (b) no restoring force, $b=0.5$, $k=0$; (c) constant drag coefficient, $b=0$, $k=0.5$; and (d) traditional oscillator, $b=0$, $k=0$.  In all cases, $\tau=1$, $\xi_o=1$, $\mathrm{X_s}=1$, $F=1$. 
	}
	\label{fig:simp_vel} 
\end{figure}

Figure~\ref{fig:simp_vel}(a) shows the typical evolution of $\dot{x}$ for the shape-dependent oscillator.   The dynamics resembles qualitatively the dynamics of the full filament model shown in the second row of figure~\ref{fig:single}. Specifically, after each switch, the velocity $\dot{x}$ first experiences a sharp decrease and remains small until the next switch. This is a joint effect of the internal restoring moment $-k\mathrm{s}$ due to the filament elasticity and the shape-dependent drag coefficient $\xi(\mathrm{s})$. If the restoring moment is eliminated ($k=0$),  the velocity profile changes such that it first decreases then increases (figure~\ref{fig:simp_vel}(b)), and if the drag coefficient is held constant ($b=0$), the decrease in velocity after each switch is more gradual (figure~\ref{fig:simp_vel}(c)). In the case of the standard geometric switch $b=k=0$, the velocity remains constant after each switch because the applied force considered here is constant (figure~\ref{fig:simp_vel}(d)).

We now consider two hydrodynamically-coupled, shape-dependent oscillators, 
\begin{equation}
\begin{split}
\xi(\mathrm{s}_i)[\dot{x}_i - v_j(x_i)]&=-k\mathrm{s}_i+\alpha_i F, \\[0.5ex]
\tau \dot{\mathrm{s}}_i &= -\mathrm{s}_i+\alpha_i F.
\label{eq:2oscillators}
\end{split}
\end{equation}
Subscripts $i,j = 1,2$ are the oscillator indices and $v_j(x_i)={\dot{x}_j}/{|x_i-x_j|}$ (with $ i\ne j$) is the  far-field approximation of the flow velocity generated by the motion of oscillator $j$ at $x_i$.
The state variable $\alpha_1$ switches between $\{-1,1\}$ once $|x_1|=\mathrm{X_s}$, while $\alpha_2$ switches between $\{-1,1\}$ once $|x_2-d| =\mathrm{X_s}$, where $d$ is the separation distance between the centers of the two oscillator trajectories.

\begin{figure}
	\centerline{\includegraphics[width=\linewidth]{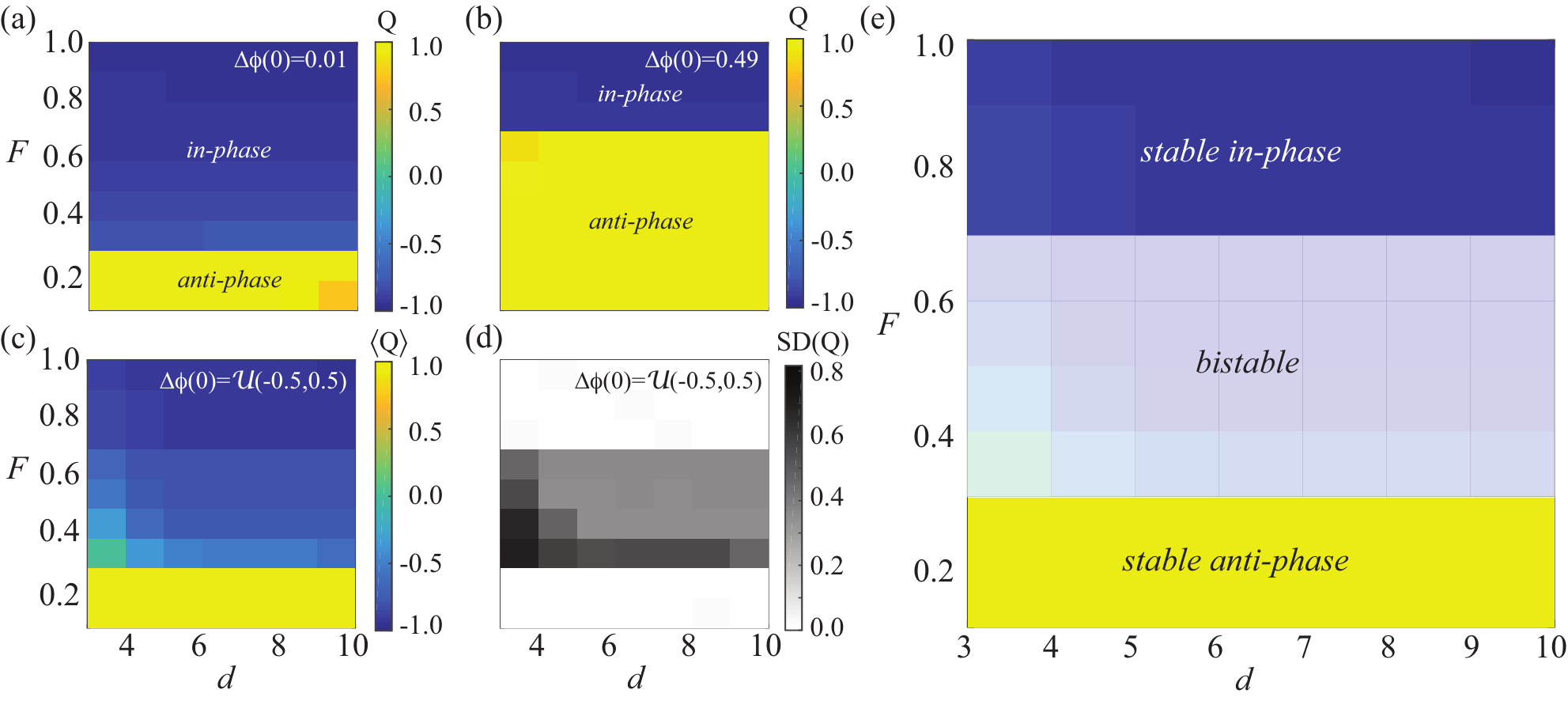}}
	\caption[]{
	Synchronization modes of  two shape dependent oscillators as a function of applied force $F$ and separation distance $d$ for
(a) nearly in-phase ($\Delta\phi(0)=0.01$), (b) nearly anti-phase ($\Delta\phi(0)=0.49$), and (c) and (d) 20 randomly chosen initial conditions.  $\langle Q\rangle$ and  $\mathrm{SD}(Q)$ are the mean and standard deviation of synchronization order parameter.  High $\mathrm{SD}(Q)$ indicates sensitivity to initial conditions.
	(e)  In-phase, anti-phase and bistable regions obtained by overlaying (c) and (d).
	In all simulations, $\tau = 1$, $b=2$, $k = 0.1$, $\mathrm{X_S}=1$, $\xi_o=1$.}
	\label{fig:simp_overlay}
\end{figure}

If $b=k=0$, the first equation in~\eqref{eq:2oscillators}  is consistent with the geometric switch oscillators in~\cite{Kotar2010,Bruot2012} with one major distinction: here the applied force has constant magnitude. In their model, the applied force depends on the particle position, which determines the type of synchronization: the two oscillators synchronize anti-phase if the force magnitude decreases as the oscillator approaches $\mathrm{X}_s$ and in-phase if the force magnitude increases. For constant force, the two oscillators do not synchronize. Nowhere bistable synchronizations are observed. Consistent with their findings, when the shape changes are not accounted for ($b=k=0$), the model in~\eqref{eq:2oscillators} exhibits no 
synchronization -- 
the two oscillators maintain their initial phase difference for all time. However, when shape changes are considered ($b\neq 0$ and $k\neq 0$), multiple synchronization modes can arise depending on the parameter values and initial conditions as shown in figure~\ref{fig:simp_overlay}.

Figures~\ref{fig:simp_overlay}(a) and (b) show the synchronization order parameter $Q$ as a function of the force magnitude $F$ and separation distance $d$ for two sets of initial conditions  $\Delta\phi(0)=0.01$ and $\Delta\phi(0)=0.49$, respectively. For both initial conditions,  in-phase synchronization is favored as $F$ increases,  albeit for different values of $F$. This tendency to synchronize in-phase at larger $F$ is consistent with the trend observed in the full filament model. Figures~\ref{fig:simp_overlay}(c) and (d) show the results of 20 Monte Carlo simulations with random initial conditions taken from a uniform distribution function: a stable anti-phase region is observed for $F<0.3$, a stable in-phase region for $F>0.7$, and a bistable region for $0.3<F<0.7$. The three regions are illustrated in figure~\ref{fig:simp_overlay}(e). 

The simple model in~\eqref{eq:2oscillators} captures some of the main features of the full filament model. In particular, it shows the presence of regions where  in-phase and anti-phase oscillations are both stable, depending on initial conditions.   This bistability is the product of the coupling between hydrodynamic interactions and shape changes. In fact, if the restoring force due to elasticity is eliminated ($k=0$ but $b>0$), the oscillators always synchronize in phase. Meanwhile, if the  dependence of drag on shape is eliminated ($b=0$ but $k>0$), the oscillators always synchronize anti-phase. The two types of synchronization modes are observed only when the two effects of $k$ and $b$ are present.
These findings imply  that the  two different synchronization modes observed in the simplified shape-dependent oscillators and in the full filament model are due to the interplay between elasticity, shape-dependent drag, and hydrodynamic coupling.

\section{Discussion}
The main contributions of this work can be summarized as follows:
\begin{itemize}
\item[] (i) We proposed a model for elastic microfilaments of finite length submerged in viscous fluid; the filaments are attached to a wall  and driven at their base by a bending moment that is geometrically-triggered to switch direction as the filament approaches  pre-defined target angles. 
\item[] (ii) We considered full hydrodynamic interactions among and within filaments. Isolated filaments were shown to undergo long-term periodic deformations that are insensitive to initial conditions and whose waviness and frequency increased with increasing the intensity of the driving moment. 
\item[] (iii) \textcolor{black}{Pairs of} filaments exhibit stable in-phase and anti-phase synchrony that are robust to initial perturbations; more interestingly, both in-phase and anti-phase synchronizations stably co-exist in regions of the parameter space (driving moment versus separation distance), with in-phase synchrony associated with higher oscillation frequencies.  These multiple synchronization modes are inherently non-linear and cannot be captured in a linear stability analysis. 
\item[] (iv) To explain the main mechanisms underlying the observed behavior, we proposed a low-order model of an elastic ``particle" that accounts for shape changes in terms of a ``lumped" shape variable that is coupled to the particle's position. The simpler model recapitulates the behavior observed in single and \textcolor{black}{pair of} filaments and  highlights the role of each component  -- elasticity, shape-dependent drag, and hydrodynamic coupling -- in the emergent behavior. 
\end{itemize}

Our low-order model is consistent with the geometric switch oscillators of~\cite{Kotar2010,Bruot2012}. In the latter,  the driving force depends on the particle configuration, and its functional form determines the type of synchronization: two oscillators synchronize anti-phase if the magnitude of the driving force decreases as each oscillator approaches its switching positions and in-phase if the force magnitude increases. Shifting between different synchronization modes requires changing the model of the driving force. In contrast, in our models, the magnitude  of the drive is independent of configuration. Stable in-phase or anti-phase as well as bistable synchronization modes, all arise 
 without the need to change the functional form of the drive. 
A transition from in-phase to anti-phase synchrony can be induced either  by varying the drive level or by perturbing the initial conditions at the same drive level.  
On a more abstract level, the dynamics in our models can be thought of as associated with one potential landscape with multiple local minima that can be visited by either changing the parameter values or the initial conditions.

These findings -- namely, the co-existence of  in-phase and anti-phase synchrony and the fact that in-phase synchrony is associated with higher frequencies and filament waviness (traveling-wave deformations) -- are consistent with experimental observations in a \textit{Chlamydomonas} biflagellate~\cite{Leptos2013}. Flagella were shown to switch stochastically between  anti-phase and in-phase states, and that the latter has a distinct waveform and significantly higher frequency {(the notation in-phase and anti-phase is reversed in~\cite{Leptos2013}).
In the context of our model, such switching could occur due to random perturbations or by varying the intensity of the internal drive. 
This is in contrast to alternating between different models of the drive characterized by different modes of synchrony~\cite{Leptos2013}. The distinction between these two views -- keeping the same form of the drive or alternating between different drive forms --  is fundamentally linked to admissible hypotheses on the physiological and biophysical mechanisms underlying
the transition between different synchronization modes. For example, in light of our results, it is plausible that transitions in biflagellar synchrony are triggered purely mechanically, say by random noise in the medium, without biochemical changes that alter the driving forces, or physiologically by modifying either the intensity of the drive or the compliance of the flagella, without inducing new behavior in the internal machinery.

\textcolor{black}{In the alga biflagellate \textit{Chlamydomonas}, mechanical coupling at the flagella base  could be playing a role in flagellar synchronization~\cite{Friedrich2012, Geyer2013,Quaranta2015, Wan2016}. Importantly, in-phase and anti-phase synchrony is also observed between a pair of pipette-held flagella of \textit{Volvox} somatic cells, where the coupling is purely hydrodynamic~\citep[supplementary movies 2 and 3]{Wan2016}.} 

\textcolor{black}{The model presented here serves to demonstrate that the interplay between elasticity, hydrodynamics and geometry-dependent actuation could give rise to multiple synchronization modes.  While in itself, the model is not intended to faithfully describe biological cilia and flagella, the outcomes of the model could serve to guide future research} 
and formulate new hypotheses regarding the mechanisms that drive and alter synchrony in biological and physical systems. 
\textcolor{black}{For example, it would be interesting to revisit the pipette-held flagella of \textit{Volvox} somatic cells and conduct systematic experiments for identifying the physical parameters leading to in-phase and anti-phase synchrony. It would also be interesting to develop experimental protocols for gradually increasing the activity of the molecular motors in reactivated axonemes as in~\cite{Geyer2016} to gauge the effect of the actuation level on synchrony.
From the modeling standpoint,} future extensions of this work will account for more accurate models of the internal driving moments (\cite{Goldstein2016b, Sartori2016}), three-dimensional filament deformations with torsion and twist~\cite{Olson2013, Man2016}, and multiple interacting filaments with application to metachronal coordination of cilia~\cite{Gueron1999, Mitran2007, Yang2008, Guo2014,Guo2016}.  \textcolor{black}{Meanwhile, we are working on including mechanical coupling at the base of the filaments to emulate basal coupling in the ciliated cell and investigate its role in filament synchronization, in the presence and absence of fluid coupling. The low-order model also presents a rich framework in which to explore synchronization of multi-particle oscillators as in~\cite{Vilfan2006, Uchida2011}.}

\section*{Acknowledgment}  The work of  H. Guo and E. Kanso is partially supported by the NSF INSPIRE grant 170731 (to E. Kanso).

\bibliographystyle{jfm}
\bibliography{reference}

\end{document}